# Experimental Setup for Studying Guiding of Proton Microbeam

G.U.L. Nagy, I. Rajta, R.J. Bereczky, and K. Tőkési

*Institute of Nuclear Research of the Hungarian Academy of Sciences (MTA Atomki), Debrecen 4001, Hungary, EU*

**Abstract.** We present the design and construction of our experimental setup for studying the transmission of proton microbeam through a single, cylindrical shape, macroscopic insulating capillary. The intensity as a function of time, the energy distribution as a function of the transmission and the deflection of the transmitted particles can be measured with the new setup.

## INTRODUCTION

Charged particles, keeping their initial charge states, can be transmitted through an insulating capillary even if the capillary axis is tilted with respect to the incident beam axis larger than the geometrical limit [1-12]. This phenomenon is called charged particle guiding. In the past few years, since the discovery of the guiding effect, a number of experimental as well as theoretical works have been published on various insulating foils like Polyethylene-Terephthalate (PET), silicon dioxide ($SiO_2$) and aluminum oxide ($Al_2O_3$) nanocapillaries with aspect ratios around 100, using slow highly charged ions.

Although during the past few years many research groups joined to this field of research and carried out various experiments with insulator capillaries many details of the interactions remained unknown. Initially insulating foils with randomly distributed nanocapillaries were used, which foreshadow many uncertainties both in experimental and theoretical points of views. For example, it is not possible to ensure a perfect parallelism of the nanocapillaries in the foil. Another significant problem is that the collective effect of all the neighboring tubes has to be taken into account for the accurate simulation.

As a new feature and to avoid these difficulties, we want to use the combination of single charged projectiles and single capillary, and we plan systematic measurements on the guiding of protons through a single, cylindrical shape macrocapillary. The other uniqueness of our work is the use of the small and well defined size of the proton beam in order of the micrometer size. With the proton microbeam, the investigation of the local charge-up inside the capillary also becomes possible.

In our forthcoming experiments we would like to try to determine whether the guiding-effect is still observable for 1 MeV proton beam. We note that the velocity of the 1 MeV protons is about 100 times larger than that of the highly charged ions used so far in the previous investigations. We plan to measure the energy distributions of proton microbeam transmission through a single macroscopic Teflon capillary as a function of time. In this work we present the design and construction of our experimental setup which makes these measurements possible.

## EXPERIMENTAL SETUP

To perform the capillary experiments with proton microbeam, the Oxford-type nuclear microprobe installed on the 0° beam line of the 5 MV Van de Graaff accelerator of the Atomki, Hungary, was modified. Figure 1 shows the target chamber after the modification with the accessories used for the experiment. Rutherford Backscattering (RBS) signal and the optical microscope are used for the sample positioning. A 2x2 array of PIN diodes collected the backscattered particles. Because of its large area (and solid angle) the efficiency of the collection was high and thus the positioning was very quick. The deflection plates are installed for determining the charge state of the transmitted beam. A commercially available lightbulb without glass housing is used as a thermal electron source at the beginning of the measurements to ensure neutral capillary inner surface.

The capillary is aligned with respect to the beam axis and the focal point with high accuracy by a high resolution, 5-axis goniometer. In our recent case the sample can be moved by 2 μm step resolution in 3 directions and can be rotated by 0.01° angular resolution around 2 axes (vertical and horizontal). These values are sufficient considering the sample dimensions (800 μm inner diameter and with aspect ratio of 55). To measure the incoming current a beam chopper placed in front of the sample is implemented. The particles backscattered from the chopper are detected with a particle detector by periodically interrupting the beam by a rotating vane. This allows us to control the incident beam current according to the particle count rate. The transmitted current is monitored by a Faraday-cup at the outlet of the capillary. The signal from the beam chopper and Faraday-cup is recorded in every second by a PC data acquisition system.

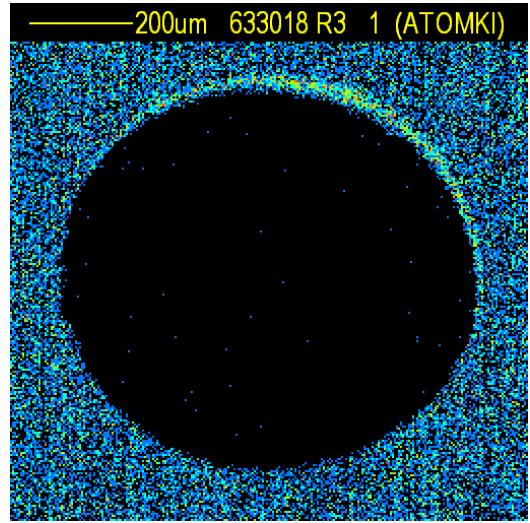

**FIGURE 2.** Sample positioning by Rutherford Backscattering mapping.

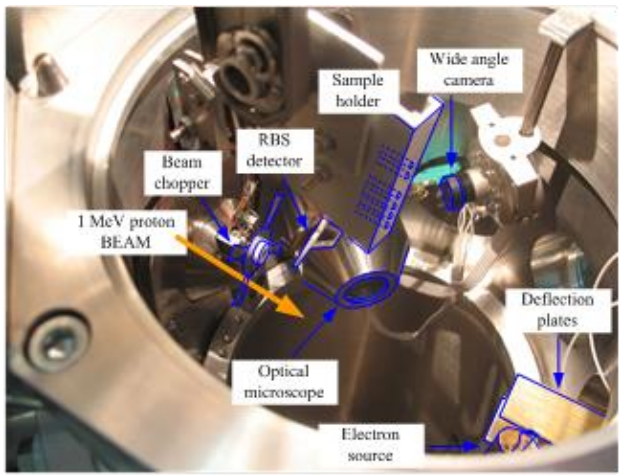

**FIGURE 1.** Photo of the target chamber with the accessories used for the capillary experiment.

The energy distribution of the transmitted particles is measured by a Hamamatsu S1223 type PIN diode, that is mounted on a rotating disk along with the Faraday-cup. This detector collects the transmitted particles at the outlet of the capillary. Since it requires low particle count rate, the incoming current was considerably reduced by closing the object collimators during the determination of the energy.

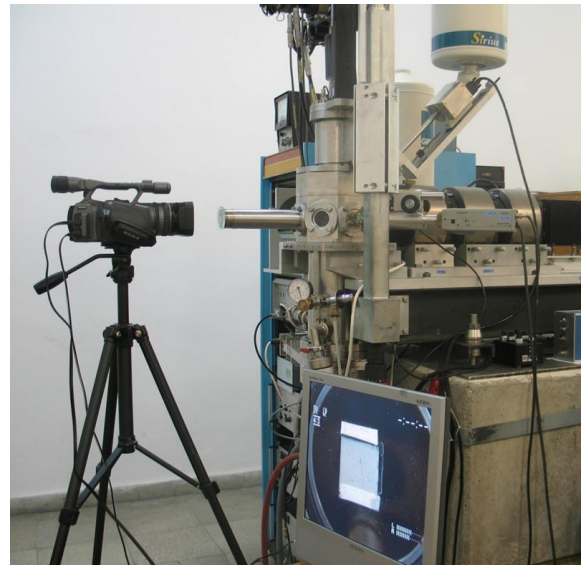

**FIGURE 3.** Photo of the microbeam setup for measuring the beam deflection and charge state.

The sample could be accurately positioned by RBS mapping technique. First the capillary was put roughly into the right place by observing it with an optical microscope. Then the beam was scanned over the capillary inlet, and an RBS map was collected. The capillary was moved until it got into the center of the scanned area (see Figure 2). After this the scanning was turned off and the beam was steered to the center of the inlet by the beam controlling software. This method can accurately position the sample.

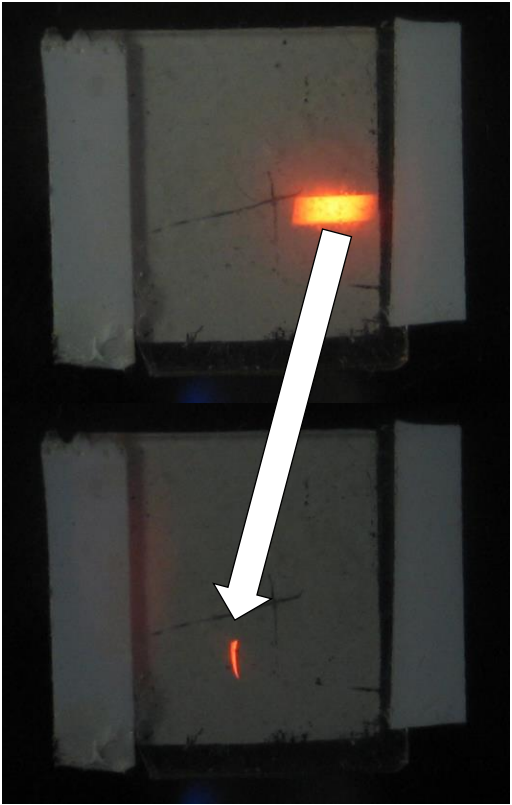

**FIGURE 4.** Photo of the fluorescent screen. The top figure is made at 0° tilt angle, while the bottom is at 1°.

For the visualization of the deflection of the transmitted beam a fluorescent screen, which has orange color when the beam reaches it, was placed at the end of the chamber. Since we plan to carry out our investigations at small angles, a 21 cm long drift tube was placed between the sample and the screen to magnify the deflection. The screen was mounted onto a transparent substrate, thus one could observe the actual position of the beam from behind. Since the sample was rotated around the vertical axis, the deflection was observable in the horizontal direction. Charge state was determined with the help of a parallel-plate deflector. 2.7 kV high voltage was applied between two parallel copper plates with 20 mm distance from each other to separate the charged and the neutralized particles. It was mounted in such a way that it resulted in a vertical deflection of the charged particles. This way we could unambiguously separate the charged particles even if the beam was horizontally deflected.

Figure 3 shows the setup for measuring the beam deflection and charge state. A high resolution digital video camera recorded the beam on the fluorescent screen. This allowed us to observe the change of the position and the shape of the beam and to determine accurately its deflection.

Figure 4 shows a snapshot of the beam on the screen. The top figure is made at 0° tilt angle, while the bottom is at 1°. The difference between the two positions of the beam is measured to be 4 mm. Considering the drift space between the target and the fluorescent screen this corresponds to 1° tilt angle, i.e. the output beam is parallel to the capillary axis.

## CONCLUSION

Experimental setup for the investigation of the ion guiding of the proton microbeam was successfully constructed. The intensity, the energy distribution, and the deflection of the transmitted particles could be determined. The charge state of the transmitted beam can be also identified. We believe that the forthcoming investigations of the proton microbeam transmission will help to reach a better understanding of the guiding phenomenon. Work along this line is in progress, and detailed experimental results will be published soon.

## ACKNOWLEDGMENTS

This work supported by the Hungarian Scientific Research Fund OTKA No. NN 103279.